\documentclass[aps,pr,twocolumn,groupedaddress,notitlepage,showpacs,floatfix,superscriptaddress]{revtex4-1}
\pdfoutput=1
\usepackage{graphicx,graphics,epsfig,subfigure,times,bm,bbm,amssymb,amsmath,amsthm,mathrsfs,MnSymbol}
\usepackage{gensymb}
\usepackage{amsfonts}
\usepackage{float}
\usepackage[matrix,frame,arrow]{xypic}
\usepackage[pdfstartview=FitH]{hyperref}
\hypersetup{
    colorlinks=true,       
    linkcolor=red,          
    citecolor=blue,        
    filecolor=magenta,      
    urlcolor=blue,           
    runcolor=cyan
}
\usepackage[pdftex]{color}
\usepackage{braket}  
\usepackage{enumerate}
\usepackage[normalem]{ulem}
\usepackage[usenames,dvipsnames]{xcolor}
\usepackage{multirow}
\usepackage{mathtools}
\usepackage{bbm}

\definecolor{orange}{rgb}{1,0.5,0}

\newcommand{\RNum}[1]{\uppercase\expandafter{\romannumeral #1\relax}}

\newcommand{\ignore}[1]{}
\usepackage{geometry}

\ignore{
\documentclass[eprintnumbers,amsmath,amssymb,onecolumn,a4paper,caption ]{article}
\usepackage{amsfonts}
\usepackage{amssymb}
\usepackage{mathrsfs}
\usepackage{mathbbold}
\usepackage{bbm}
\usepackage{mathrsfs}
\usepackage{dcolumn}
\usepackage{bm}
\usepackage{times,epsfig,amssymb,amsmath}
\usepackage{float}
\usepackage{subfigure}
\usepackage{geometry}\geometry{left=2.5cm,right=2.5cm,top=3cm,bottom=3cm}

\usepackage{color}

}

\begin{document}
	\title{Topological band theory for non-Hermitian systems from the Dirac equation}
	
	\author{Zi-Yong Ge}
	\affiliation{Institute of Physics, Chinese Academy of Sciences, Beijing 100190, China}
	\affiliation{School of Physical Sciences, University of Chinese Academy of Sciences, Beijing 100190, China}
	
    \author{Yu-Ran Zhang}
    \email{yrzhang@csrc.ac.cn}
    \affiliation{Beijing Computational Science Research Center, Beijing 100193, China}
	\affiliation{Theoretical Quantum Physics Laboratory, RIKEN Cluster for Pioneering Research, Wako-shi, Saitama 351-0198, Japan}

	\author{Tao Liu}
	\email{tao.liu@riken.jp}
	\affiliation{Theoretical Quantum Physics Laboratory, RIKEN Cluster for Pioneering Research, Wako-shi, Saitama 351-0198, Japan}
	
	 \author{Si-Wen Li}
	 \affiliation{Department of physics, Dalian Maritime University, Dalian 116026, China}

    \author{Heng~Fan}
    \email{hfan@iphy.ac.cn}
    \affiliation{Institute of Physics, Chinese Academy of Sciences, Beijing 100190, China}
    \affiliation{School of Physical Sciences, University of Chinese Academy of Sciences, Beijing 100190, China}
    \affiliation{CAS Center for Excellence in Topological Quantum Computation, UCAS, Beijing 100190, China}
    \affiliation{Songshan Lake Materials Laboratory, Dongguan 523808, China}

    \author{Franco Nori}
    \affiliation{Theoretical Quantum Physics Laboratory, RIKEN Cluster for Pioneering Research, Wako-shi, Saitama 351-0198, Japan}
	\affiliation{Department of Physics, University of Michigan, Ann Arbor, Michigan 48109-1040, USA}

\begin{abstract}
We identify and investigate two classes of non-Hermitian systems, i.e., one resulting from Lorentz-symmetry violation (LSV) and the other from a complex mass (CM) with Lorentz invariance.
The mechanisms to break, and approaches to restore, the bulk-boundary correspondence in these two types of non-Hermitian systems are clarified. The non-Hermitian system with LSV shows a non-Hermitian skin effect, and its topological phase can be characterized by mapping it to the Hermitian system via a non-compact $U(1)$ gauge transformation. In contrast, there exists no non-Hermitian skin effect for the non-Hermitian system with CM. Moreover, the conventional bulk-boundary correspondence holds in this (CM) system. We also consider a general non-Hermitian system in the presence of both LSV and CM, and we generalize its bulk-boundary correspondence.
\end{abstract}

\maketitle

\section{Introduction}%
Topological band theory for Hermitian systems, such as topological insulators and topological superconductors \cite{Thouless,Read,Kitaev1,Kane1,Kane2,Moore,Fu1,Fu2,Schnyder,Kitaev2,Hasan,Qi,Bansil,Chiu}, has made remarkable progress in the past two decades.
Recently, topological phases of non-Hermitian systems have been theoretically studied
in  topological insulators \cite{Esaki,Hu,Yuce,Klett,Lieu,Yin,Takata2018,Kawabata2018,Liu,Jin2019,Edvardsson2019,Lee201919,Alvarez1,Yoshida2},
topological superconductors \cite{Wang,Menke,Li,Kawabata,Kawabata2019},
Weyl semimetals \cite{Gonzlez,Cerjan,Budich2019},
etc. \cite{Rudner,JGong,Liang,Leykam,Gong1,WHu,Rivolta,Shen1,CHHua2019,Kawabata1,zeng2019,Konstantin1,Konstantin2,Kawabata20192,HZhou1,HZhou2,Borgnia,Yoshida1},
and have been experimentally observed in Refs.~\cite{Zeuner,Zhan,Weimann,Xiao,Parto,Zhu}.
Non-Hermitian Hamiltonians can describe open systems with gain and/or loss \cite{Rotter}, interacting and disordered systems with finite lifetimes of quasiparticles \cite{Zyuzin,Kozii,Papaj,Shen2}, and have many unique features, e.g., the existence of exceptional points causing eigenstates to coalesce and making Hamiltonians non-diagonalizable \cite{Bender,Berry,Heiss,Moiseyev,Ozdemir2019,Ganainy}. Non-Hermitian Hamiltonians have been successfully applied to explain experiments in various platforms including,  photonic \cite{Makris,Chong,Regensburger,Jing,Hodaei,Peng1,Feng,Peng2,Liu1,Kawabata3,Lv,Ashida,Zhang} and mechanical systems~\cite{Jing1, Jing2, Liu12017}. The introduction of non-Hermiticity into band theory induces many novel topological properties, which are significantly different from their Hermitian counterparts, e.g., Weyl exceptional rings \cite{Xu}, bulk Fermi arcs \cite{Zhou} and breakdown of conventional bulk-boundary correspondence \cite{Lee,Yao,Xiong,Yao2}.  In contrast to the Hermitian case, the bulk spectra of non-Hermitian systems strongly rely on the boundary conditions \cite{Kunst,Yao,Xiong,Yao2}. Therefore, topological invariants, defined by  non-Hermitian Bloch Hamiltonians, usually fail to  characterize topological phases in non-Hermitian systems, which leads to the breakdown of the conventional bulk-boundary correspondence.  Although many efforts have been made to propose new topological invariants, e.g. non-Bloch winding and Chern numbers \cite{Yao, Yao2}, to restore the bulk-boundary correspondence of non-Hermitian Hamiltonians \cite{Kunst, Yao, Yao2, Zirnstein2019, Herviou2019}, it remains a challenge to understand and characterize the topological phases of non-Hermitian systems. For example, for a non-Hermitian Hamiltonian, should its topological phase follow the Block-wave or non-Bloch-wave behavior? This is not uncovered in Refs.~\cite{Yao, Yao2}.

In this paper, we investigate the topological phases of non-Hermitian systems.
According to both Dirac and current-conservation equations, non-Hermitian systems can mainly be classified into two classes: one resulting from Lorentz symmetry violation (LSV), and the other from a complex mass (CM) with Lorentz invariance. We clarify the mechanisms to break the conventional bulk-boundary correspondence in these two types of non-Hermitian systems, and develop approaches to generalize the bulk-boundary correspondence. In particular, the topological phases of non-Hermitian Hamiltonians with LSV can be described by non-Bloch topological invariants, while those with CM follow the Bloch-wave behavior. The non-Hermitian Su-Schrieffer-Heeger (SSH) \cite{Su}, Qi-Wu-Zhang (QWZ) \cite{Qi2} models and the disordered Kitaev chain \cite{Kitaev1} exemplify our approaches. Remarkably, our approach can unperturbatively predict the topological phases of 2D non-Hermitian systems. We also discuss a general non-Hermitian system containing \textit{both} LSV and CM non-Hermiticities.

The remaining of this paper is organized as follows. In Sec.~\ref{Hermitian Dirac equation}, we give a short review of Hermitian Dirac equation. In Sec.~\ref{Non-Hermiticity with LSV}, the LSV non-Hermitian systems are studied.
In Sec.~\ref{Non-Hermiticity with CM}, we discuss the  non-Hermitian systems with CM.
Non-Hermiticity with mixed LSV and CM is discussed in Sec.~\ref{Non-Hermiticity with mixed LSV and CM}.
Finally, we summarize the results of this paper in Sec.~\ref{Conclusion}.

\section{Hermitian Dirac equation}\label{Hermitian Dirac equation}
Generally, topological band systems can be described by the Dirac equation~\cite{Qi, Shen}.
The Dirac Hamiltonian reads
\begin{eqnarray}\label{de}
 \mathcal{H} = \bar{\psi}(i\vec{\nabla}\cdot\vec{\gamma}+m)\psi,
 \end{eqnarray}
where $m$ is the mass of the fermion, $\psi$ is the field operator of spinors, $\bar{\psi}=\psi^\dagger\gamma^0$, and $\gamma^\mu$s satisfy the Clifford algebra $\{\gamma^\mu,\gamma^\nu\}=2\eta^{\mu\nu}$, with $\eta^{\mu\nu}=\mathrm{diag}\{ 1,-1,-1,-1\} $ being the Minkowski metric tensor. In addition, we have the conserved current equation
\begin{eqnarray}
\partial_t\rho-\vec{\nabla} \cdot \vec{j}=0,
\end{eqnarray}
where $\rho = \bar{\psi}\gamma^0\psi$ and $\vec{j} = \bar{\psi}\vec{\gamma}\psi$.
In $k$ space, the spectrum of Eq.~(\ref{de}) is
\begin{eqnarray}
E_{\pm}^{\text{DE}}(k) = \pm\sqrt{k^2+m^2}.
\end{eqnarray}
For $m=0$, there exists zero-energy states, representing the critical points of the topological quantum phase transition. If we consider a domain-wall defect sandwiched by two regions with opposite signs of the mass, there exist gapless boundary modes localized at the interface. In the following sections, we generalize the Hermitian Dirac and current equations to the non-Hermitian cases.

\section{Non-Hermiticity with LSV}\label{Non-Hermiticity with LSV}

In this section, we introduce the first class of non-Hermitian systems, i.e, non-Hermiticity with LSV.
We begin with the LSV Dirac equation and then generalize it to the lattice models.
We find that the conventional bulk-boundary correspondence of these LSV systems breaks down due to the intrinsic currents,
and the non-Hermitian skin effect emerges \cite{Yao,Yao2}.

\subsection{Lorentz-symmetry-violation Dirac equation}
As $\mathcal{H}$ is $U(1)$ gauge invariant, we only need to consider a compact $U(1)$ gauge group in the Hermitian case. With a specific non-compact $U(1)$ gauge transformation on $\mathcal{H}$, then
\begin{eqnarray} \label{lvl}
 \mathcal{H}_{\text{LV}} = \bar{\psi}[\vec{\gamma}\cdot(i\vec{\nabla}+\vec{A})+m]\psi,
\end{eqnarray}
where $\vec{A} = \nabla\chi =\vec{A}^{\textrm{r}}+i\vec{A}^{\textrm{i}}$ is a complex vector potential, with real vectors
$\vec{A}^{\textrm{r}}$ and $\vec{A}^{\textrm{i}}$.
Here, the non-compact $U(1)$ gauge transformation $e^{S}$ on spinors satisfies 
\begin{eqnarray}
\psi \rightarrow e^{S}\psi, \ \ \ \bar{\psi} \rightarrow \bar{\psi}e^{-S}.
\end{eqnarray}
In this case, $\mathcal{H}_{\text{LV}}$ becomes non-Hermitian for
$\vec{A}^{\textrm{i}} \neq 0$. Here, the Lorentz symmetry is broken, the compact part of $U(1)$ gauge symmetry is
not broken \cite{Qi2008}, and we label this type of non-Hermiticity as LSV
\cite{Alexandre}.  The eigenenergies of $\mathcal{H}_{\text{LV}}$ are real, and have the same values as those of the
Hermitian Hamiltonian $\mathcal{H}$ in spite of the non-Hermitian terms. However, in contrast to the eigenstates of
$\mathcal{H}$, the real-space right (left) eigenstates of the continuum non-Hermitian Hamiltonian $\mathcal{H}_{\text{LV}}$
have an extra phase $e^{-i\chi}$ ($e^{i\chi}$) with complex $ \chi $. Therefore, the states of $\mathcal{H}_{\text{LV}}$
become exponentially localized, which is exactly the non-Hermitian skin effect \cite{Yao,Yao2}. Furthermore, because
$\mathcal{H}$ and $\mathcal{H}_{\text{LV}}$ can be transformed to each other by a gauge transformation, they are
topologically equivalent.

For non-Hermitian systems, the real-space bands can be considerably different from those in  $k$ space. The eigenenergies of $\mathcal{H}_{\text{LV}}$ with a constant $\vec{A}$ in $k$ space are
\begin{eqnarray}
E_{\pm}^{\text{LSV}}(k) = \pm\sqrt{(\vec{k}+\vec{A}^{\textrm{r}}+i\vec{A}^{\textrm{i}})^2+m^2},
\end{eqnarray}
and can be complex, while they are real in real space. Therefore, the energy spectra of  non-Hermitian systems strongly depend on its boundary conditions. Moreover, the bulk-boundary correspondence in this non-Hermitian system with LSV is correspondingly broken.

To uncover the nature of such breakdown, we consider the current equation for $\mathcal{H}_{\text{LV}}$~\cite{Alexandre}:
\begin{eqnarray} \label{cce2}
\partial_t\rho-\vec{\nabla}\cdot\vec{j}= -2\vec{A}^{\textrm{i}}\cdot\vec{j}.
\end{eqnarray}
The right hand of Eq.~(\ref{cce2}) is a source term, making the solutions of $\rho$ and $\vec{j}$ exponentially localized at the edge with the form $e^{-i\chi}$ ($e^{i\chi}$).
In addition, this source is a classical vector field $\vec{A}^{\textrm{i}}$ resulted from non-Hermiticity. To distinguish it with the one in CM case discussed in Sec.~\ref{Non-Hermiticity with CM}, we call it an intrinsic current.
The intrinsic current  suffers from an ambiguity on orientation, when going from open to periodic boundary conditions.
This leads to the breakdown of the conventional bulk-boundary correspondence in non-Hermitian system with LSV. To preserve the current conservation and overcome this ambiguity, we replace $\vec{k}$ with $\left(\vec{k}-\vec{A}\right)$  to cancel the effect of the gauge transformation. Then, after applying the Fourier transformation to have a real-space form, the conserved current equation is restored.
Moreover, according to topological band theory, the topology relies on the topological properties of the wave function.
Thus, the non-compact $U(1)$ transformation,  a local continuous mapping of the wave function, does not change the topology of the system.
Therefore, the topological invariant, defined by the $k$ space Hamiltonian after replacing $\vec{k}$ with $(\vec{k}-\vec{A})$, reflects the topological phases of $ \mathcal{H}_{\text{LV}} $ in real space, which restores the bulk-boundary correspondence.

\subsection{Lorentz-symmetry-violation lattice models}
As in the case for the above continuum model, the topological phases of the lattice model in the presence of the non-Hermiticity with LSV can be also investigated from the perspective of the non-compact $U(1)$ gauge transformation. To clarify this, we begin with a Hermitian lattice Hamiltonian
\begin{eqnarray}
\mathcal{H}_0=\sum_{\alpha\beta ij} \!c_{\alpha, i}^\dagger h_{ij}^{\alpha\beta} c_{\beta, j},
\end{eqnarray}
with lattice sites $i,j$  and band indices $\alpha,\beta$. By a non-compact $U(1)$ gauge transformation \cite{Bernevig}, $\mathcal{H}_0$ becomes
\begin{eqnarray}
\tilde{\mathcal{H}}=e^{-S}\mathcal{H}_0e^{S}=\sum_{\alpha\beta ij}c_{\alpha, i}^\dagger h_{ij}^{\alpha\beta}\exp({iA_{ij}}) c_{\beta, j} ,
\end{eqnarray}
where $A_{ij}$ is a lattice vector potential. If $e^{S}$ is unitary with a real $A_{ij}$, $\tilde{\mathcal{H}}$ remains Hermitian. However, for the non-unitary transformation $e^{S}$ with a complex $A_{ij}$, $\tilde{\mathcal{H}}$ becomes non-Hermitian with LSV. Moreover, as in the case of the continuum model, $\tilde{\mathcal{H}}$ exhibits the non-Hermitian skin effect, and breaks the conventional bulk-boundary correspondence. Because $\tilde{\mathcal{H}}$ and $\mathcal{H}_0$ are topologically equivalent, the topological phases of the Hamiltonian $\tilde{\mathcal{H}}$ in the presence of non-Hermiticity with LSV can be characterized by mapping it to the Hermitian form with a non-compact $U(1)$ gauge transformation.

\begin{figure}[t]
	\includegraphics[width=0.47\textwidth]{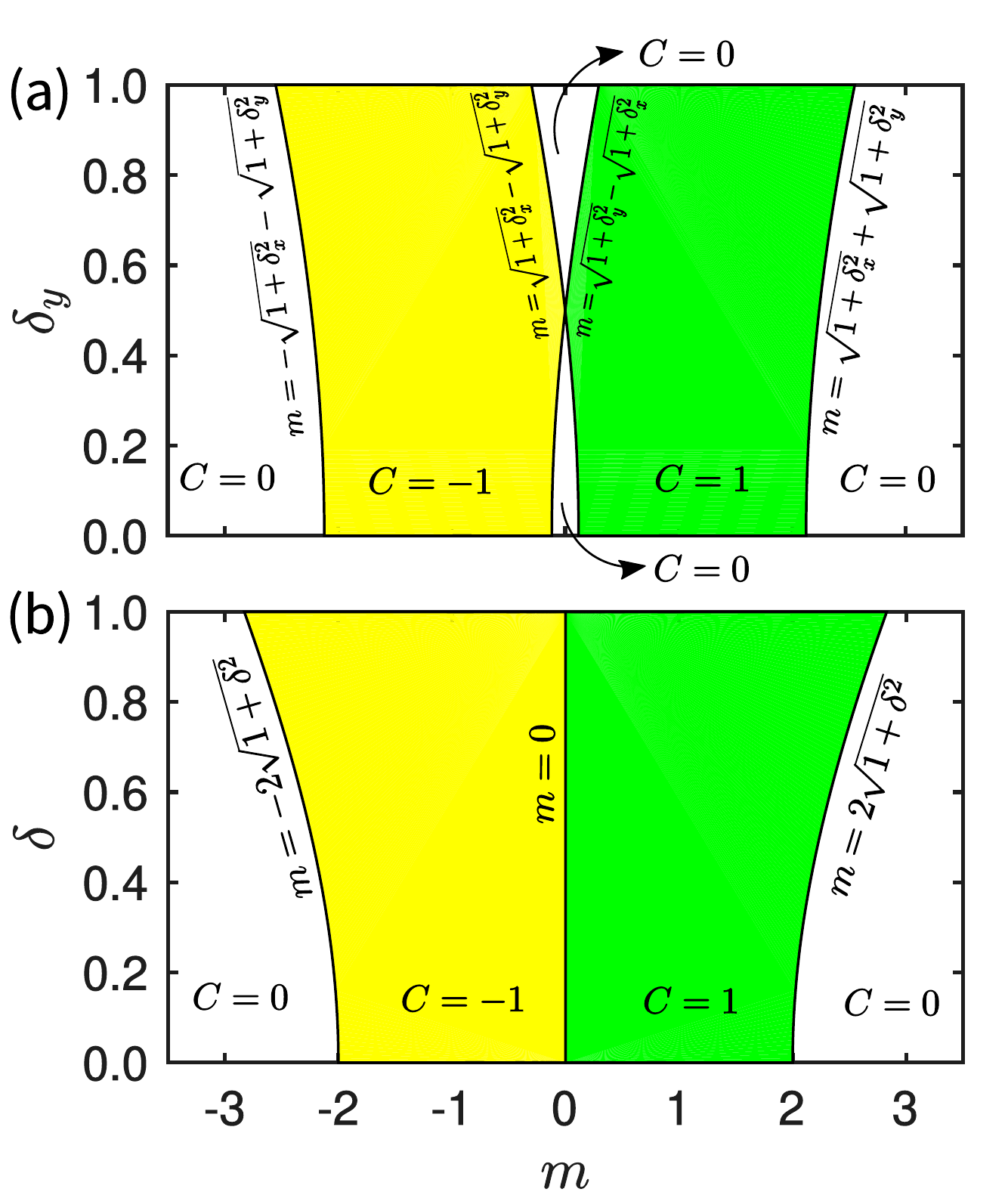}	
	\caption{Phase diagram of the 2D non-Hermitian QWZ model $H_2$, with  the corresponding Chern numbers $C$s. (a) For the general case, there are four topological phase boundaries with $m = \pm(\sqrt{1+\delta_x^2}+\sqrt{1+\delta_y^2})$, and $m = \pm(\sqrt{1+\delta_x^2}-\sqrt{1+\delta_y^2})$. In the vicinity of $m=0$, topologically trivial regions exist. Here, $\delta_x = 1/2$. (b) When $\delta_x=\delta_y=\delta$, there exist three phase boundaries with $m = \pm2\sqrt{1+\delta^2}$ and $m = 0$.}
	\label{fig_pd}
\end{figure}

\subsection{Two examples}
We now consider two concrete 1D and 2D non-Hermitian models with LSV. For the 1D SSH model with an asymmetric intercell hopping:
\begin{align} \label{h1} \nonumber
H_1 = &\sum_n t_1c_{\text A,n}^\dagger c_{\text B,n}+t_1c_{\text B,n}^\dagger c_{\text A,n} +( t_2+\Gamma)c_{\text A,n+1}^\dagger c_{\text B,n}\\
&+( t_2-\Gamma)c_{\text B,n}^\dagger c_{\text A,n+1}
\equiv\! \sum_{\alpha\beta i j} \!c_{\alpha,i}^\dagger h_{ij}^{\alpha\beta} c_{\beta,j},
\end{align}
where $\alpha,\beta$ are orbital indices representing the $A$,$B$ sub-lattices, respectively.
We rewrite $H_1$ as
\begin{eqnarray}
H_1 = \sum_{\alpha\beta ij} c_{\alpha,i}^\dagger \tilde{h}_{ij}^{\alpha\beta}\exp({iA_{ij}}) c_{\beta,j},
\end{eqnarray}
where $\tilde{h}_{ij}^{\alpha\beta}$ is the standard Hermitian SSH model.
Substituting it into Eq.~(\ref{h1}), we can obtain
\begin{eqnarray}
A_{ij}\!\!= \!\!-i\ln \sqrt{(t_2-\Gamma)/(t_2+\Gamma)}.
\end{eqnarray}
The modified intracell and intercell hopping strengths appearing in $\tilde{h}_{ij}^{\alpha\beta}$ are
\begin{eqnarray}
\tilde{t}_1=t_1  \ \  \text{and} \ \ \ \tilde{t}_2= \sqrt{(t_2-\Gamma)(t_2+\Gamma)},
\end{eqnarray}
respectively. According to the above discussions, the non-Hermitian Hamiltonian $H_1$ and the transformed Hermitian Hamiltonian
\begin{eqnarray}
\tilde{H}_1 = \sum_{\alpha\beta ij} c_{\alpha,i}^\dagger \tilde{h}_{ij}^{\alpha\beta} c_{\beta,j}
\end{eqnarray}
 are topologically equivalent. Therefore, $H_1$ is topologically nontrivial for $-\tilde{t}_2<t_1<\tilde{t}_2$. Note that this generalized bulk-boundary correspondence is the same as the one derived by the non-Bloch-wave method in Ref.~\cite{Yao}.

We then consider the 2D non-Hermitian QWZ model \cite{Yao2,Qi2}, which describes a Chern insulator:
\begin{align} \label{hc}\nonumber
H_2(\vec{k}) = &(\sin k_x +i\delta_x)\sigma_x+(\sin k_y +i\delta_y)\sigma_y\\
&+(m-\cos k_x - \cos k_y)\sigma_z.
\end{align}
To restore the bulk-boundary correspondence, the non-Bloch-wave method in perturbation theory was used in \cite{Yao2}; while we can solve it exactly. We map $H_2(\vec{k})$ to the corresponding Hermitian Hamiltonian by a non-compact $U(1)$ gauge transformation by replacing $(k_x,k_y)$ with $(k_x + iA_x,k_y+iA_y)$. Here, both $A_x$ and $A_y$ are real and $\vec{k}$-independent, and they can be solved by letting the spectrum of the corresponding Hamiltonian to be real (see Appendix A). Generally, there exist four topological phase boundaries:
\begin{eqnarray} \nonumber
m = \pm(\sqrt{1+\delta_x^2}+\sqrt{1+\delta_y^2}),\\
m = \pm(\sqrt{1+\delta_x^2}-\sqrt{1+\delta_y^2}),
\end{eqnarray}
as shown in Fig.~\ref{fig_pd}(a).
For the specific case $\delta_x=\delta_y=\delta$,  three topological phase boundaries can be found as
\begin{eqnarray} \nonumber
m = 0, \ \ \ \text{and} \ \ \ \pm2\sqrt{1+\delta^2},
\end{eqnarray}
as shown in Fig.~\ref{fig_pd}(b). In particular, for a small $\delta$, the topological phase boundaries to first-order approximation are
\begin{eqnarray}
m = \pm(2+\delta^2),
\end{eqnarray}
which are consistent with the perturbation results in Ref.~\cite{Yao2}.

 \begin{figure}[t]
 	\includegraphics[width=0.47\textwidth]{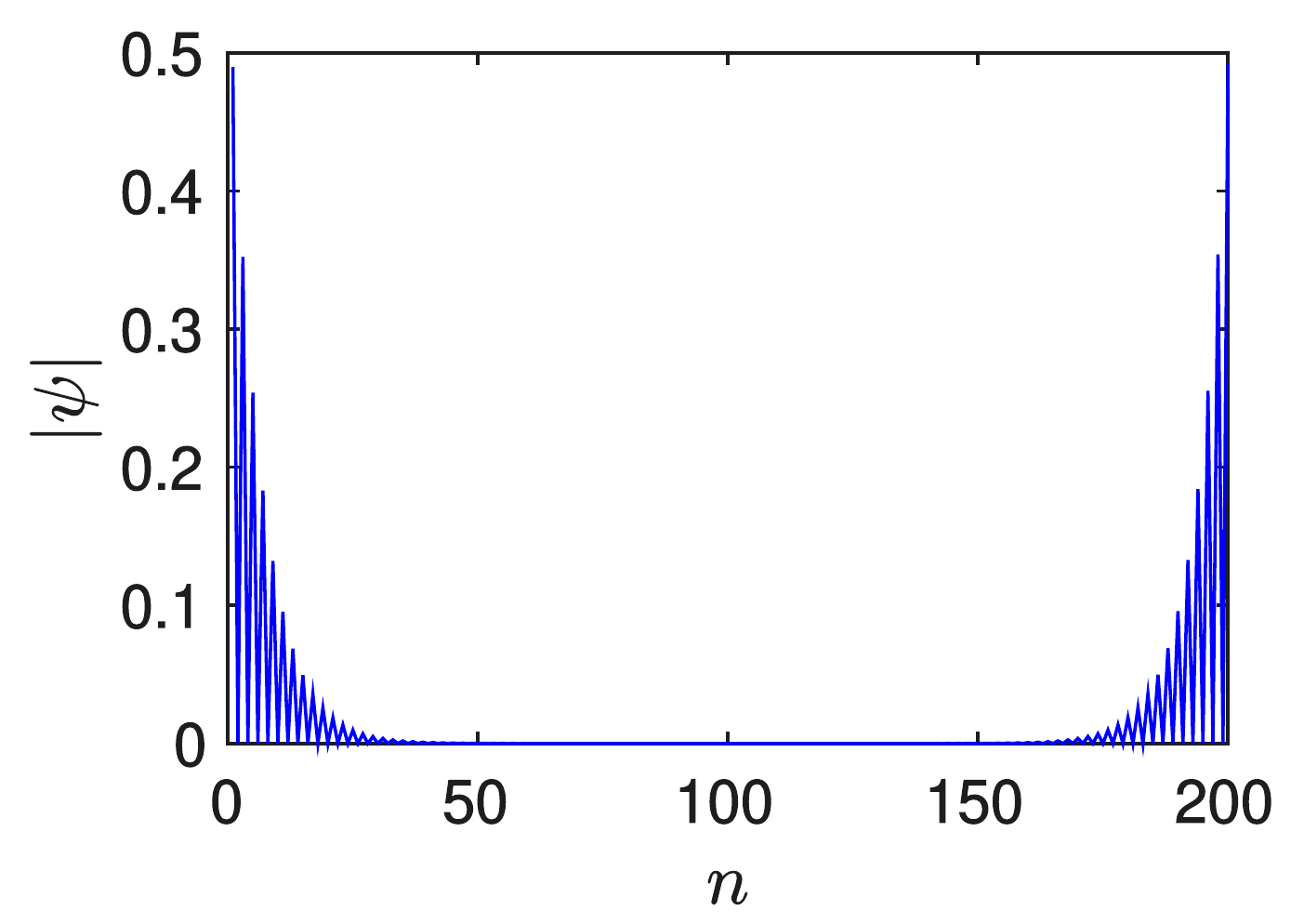}	
 	\caption{ (Color online).Spatial distribution of the zero-energy modes along the lattice with $t_1 = 0.4$. The zero-energy modes are localized at both edges, indicating there is no non-Hermitian effect.}
 	\label{h2}
 \end{figure}

  \begin{figure*}[t]
  	
  	\includegraphics[width=0.97\textwidth]{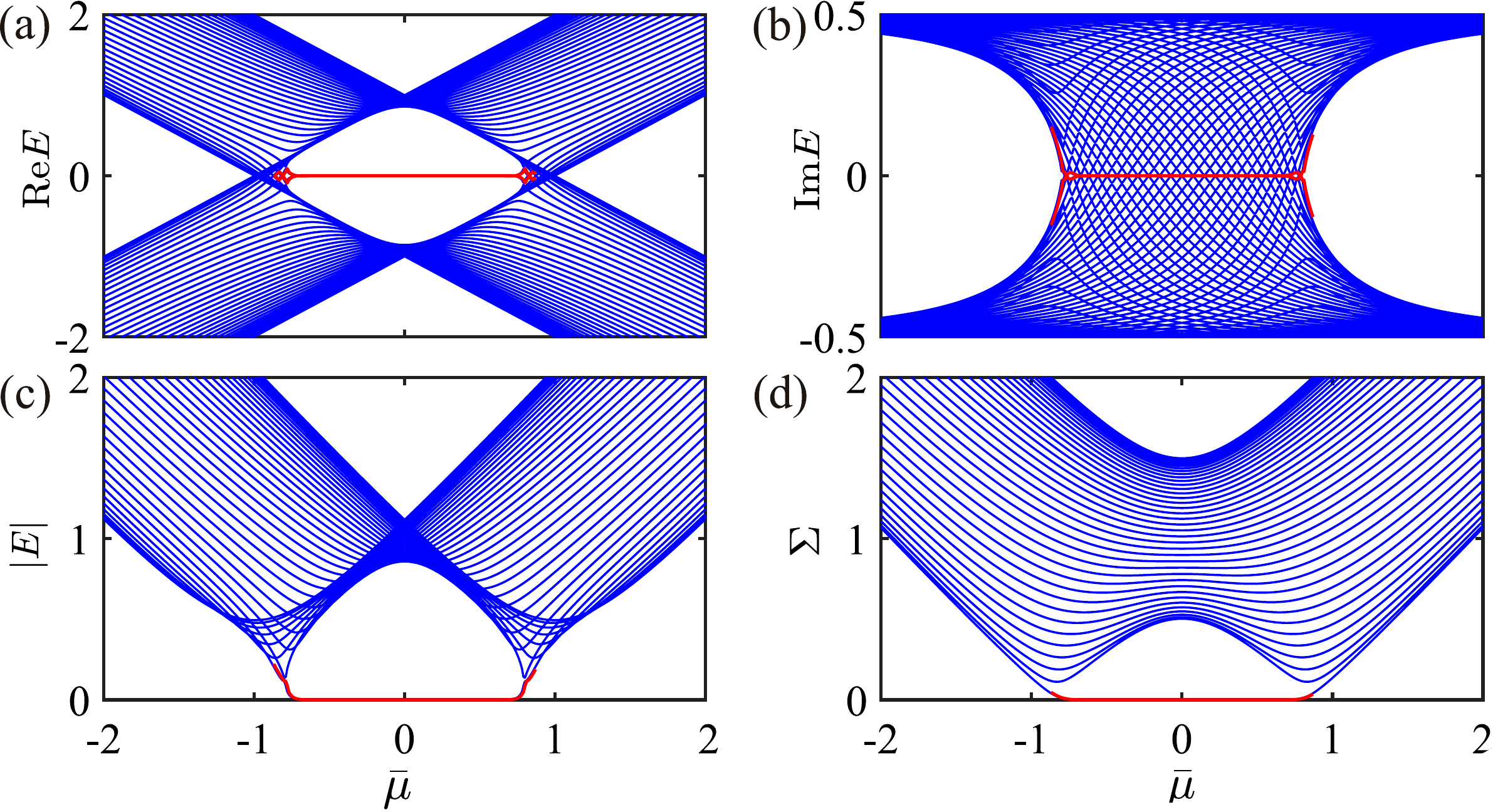}	
  	\caption{(a) Real part,  (b) imaginary part,  (c) absolute values, and (d) singular values of the energy spectra of the effective Hamiltonian $ H_{\text{eff}} (k) $ as a function of $ \bar{\mu} $ with  open boundary conditions.  For $|\bar{\mu}|<\sqrt{1-\Gamma^2}=0.87$, the system is topologically nontrivial supporting Majorana zero modes (see red curves). According to the (c,d), the singular values of $ H_{\text{eff}} (k) $ directly reflect the topological phase of the non-Hermitian systems with CM. Here, the number of unit cells is 80, with $\Gamma=0.5$, $t=\Delta=1/2$, and $V_0 = 2$.}
  	\label{fig_kc}
  \end{figure*}

\section{Non-Hermiticity with CM}\label{Non-Hermiticity with CM}
In addition to the non-Hermiticity induced by LSV, the non-Hermiticity  can also  result from generalizing some parameters from real to complex. One typical example is the non-Hermitian system whose mass term becomes complex, i.e., $m= m^\textrm{r} + i m^\textrm{i}$, with real $m^\textrm{r} $ and $m^\textrm{i}$.
In this section, we investigate the topological properties of the  non-Hermitian systems with CM. We find that there exists no non-Hermitian skin effect, and the bulk-boundary correspondence holds.

\subsection{Complex-mass Dirac equation}
The Dirac Hamiltonian of CM system reads
\begin{eqnarray} \label{dhc}
\mathcal{H}_{\text{CM}} = \bar{\psi}(i\vec{\nabla}\cdot\vec{\gamma}+m^{\textrm{r}} + im^{\textrm{i}})\psi,
\end{eqnarray}
which is  with Lorentz invariance and compact $U(1)$ gauge invariance.  In contrast to the non-Hermitian Hamiltonian with LSV, it cannot be mapped to the Hermitian form by a non-compact $U(1)$ gauge transformation. The current equation is calculated as
\begin{eqnarray} \label{cce3}
\partial_t\rho-\vec{\nabla} \cdot \vec{j}=2m^{\textrm{i}}\gamma^0\rho.
\end{eqnarray}
As in the case of LSV, the right hand of Eq.~(\ref{cce3}) is also a source term resulted from non-Hermiticity,
making the solution of $\rho$ exponentially decay or enhance in the time domain.
However, this source term in Eq.~(\ref{cce3}) is a classical scalar field and not orientable, thus there exists no intrinsic current. 
Instead, the system decays, and thus behaves like a finite-lifetime particles due to this source.
Note that the Hamiltonian Eq.~(\ref{dhc}) is different from the ones in
Refs.~\cite{Alexandre2015,JonesSmith}, where non-Hermitian mass matrices with different symmetries are discussed.

According to the zero-mode domain wall solution of $\mathcal{H}_{\text{CM}}$, $m^{\textrm{r}}$ contributes to edge localization, while $m^{\textrm{i}}$ leads to  oscillations. Therefore, $m^{\textrm{r}} =0 $ is the critical point.
In $k$ space, the dispersion relation of $\mathcal{H}_{\text{CM}}$ is
\begin{eqnarray}
E_\pm^\text{CM}(k) = \pm\sqrt{k^2+(m^{\textrm{r}}+im^{\textrm{i}})^2},
\end{eqnarray}
where the upper and lower bands coalesce at the exceptional points  $k_{\textrm{EP}}=\pm |m^\textrm{i}|$ for $m^\textrm{r}  =0$, and the topological phase boundaries are determined by $|E^\text{CM}| = 0$. In contrast to the system with LSV, since the source of non-Hermitian system with CM is not orientable, i.e., there is no intrinsic current, it will not suffer from any ambiguity on orientation, when going from open to periodic boundary conditions. Thus, the energy spectrum is not sensitive to the boundary condition. Therefore, the conventional bulk-boundary correspondence holds for the system in the presence of non-Hermiticity with CM. In Appendix B, we give a detailed geometric description of the topological phase transition for this kind of non-Hermitian system. Note that the above discussion for the non-Hermitian continuum model can be directly generalized to the non-Hermitian lattice model with CM, where there exists no non-Hermitian skin effect, and the conventional bulk-boundary correspondence holds.

 Alternatively, as disscussed in Ref.~\cite{Ueda}, we can consider a minimal coupling  to a two-level environment to describe the non-Hermitian systems with CM. The coupled Hamiltonian can be written as
 \begin{eqnarray}
 H_{cp} = \mathcal{H}_{\text{CM}}\otimes\sigma_{+}+\mathcal{H}^\dagger_{\text{CM}}\otimes\sigma_{-},
 \end{eqnarray}
 where $\sigma_{\pm}\equiv(\sigma_x\pm i\sigma_y)/2$.
 It is shown that the eigenvalues and eigenvectors of $H_{cp}$ are the singular values and singular matrices of $\mathcal{H}_{\text{CM}}$ (see Appendix C). Therefore, the minimal coupling is equivalent to solving the singular value decomposition (SVD) of $\mathcal{H}_{\text{CM}}$, where $|E^\text{CM}|=0$ represents zero-value singular values of $\mathcal{H}_{\text{CM}}$. Therefore, we can also use the SVD to explore the topological properties of the non-Hermitian systems with CM, and overcome the precision problem of numerical diagonalization of non-Hermitian Hamiltonians.

 \subsection{Complex-mass SSH model}
 For a non-Hermitian system with a complex mass, there exists no non-Hermitian skin effect, and the conventional bulk-boundary correspondence holds.  To exemplify this, we consider a non-Hermitian SSH model with complex mass. The real-space Hamiltonian reads
 \begin{eqnarray} \label{nhssh2} \nonumber
 H_3 = &&\sum_n (t_1+i\kappa)c_{\text A,n}^\dagger c_{\text B,n}+(t_1+i\kappa)c_{\text B,n}^\dagger c_{\text A,n} \\
 &&+ t_2c_{\text A,n+1}^\dagger c_{\text B,n}+t_2c_{\text B,n}^\dagger c_{\text A,n+1},
 \end{eqnarray}
 whose eigenenergies are complex. It cannot be mapped to a Hermitian model with a non-compact $U(1)$ gauge transformation. In addition, according to the real-space eigenstates, there exists no non-Hermitian skin effect. Applying the Fourier transformation, we have the $k$-space Hamiltonian as
 \begin{eqnarray}  \nonumber
 H_3(k) &&= (t_1+i\kappa+t_2\cos k)\sigma_x + t_2\sin k \sigma_y\\ [0.2cm]
 &&=\left(\begin{matrix}
 0 & t_1+i\kappa+t_2e^{-ik} \\ \nonumber
 t_1+i\kappa+t_2e^{ik} & 0
 \end{matrix}\right) \\[0.2cm]
 &&=\left(\begin{matrix}
 0 & h_1(k)\\
 h_2(k) & 0
 \end{matrix}\right).\label{nhssh2k}
 \end{eqnarray}
 The dispersion relation is
 \begin{eqnarray} \label{dr2k} \nonumber
 E_{\pm}(k)=\pm\sqrt{t_1^2+t_2^2-\kappa^2+2t_1t_2\cos k+2i\kappa (t_1+t_2\cos k)}.\\
 \end{eqnarray}
 Letting $|E_{\pm}(k)|=0$,  we can obtain the topological phase transition points as $t_1^2=t_2^2-\kappa^2$. The winding numbers of $h_1(k)$ and $h_2(k)$ that have either opposite values or simultaneously zero values.  For $t_1^2<t_2^2-\kappa^2$, the winding numbers are non-zero, indicating that the system is topologically nontrivial. Moreover, the zero-energy boundary modes are localized at both edges (see Fig.~\ref{h2}).

 \begin{figure*}[t]
 	\includegraphics[width=0.97\textwidth]{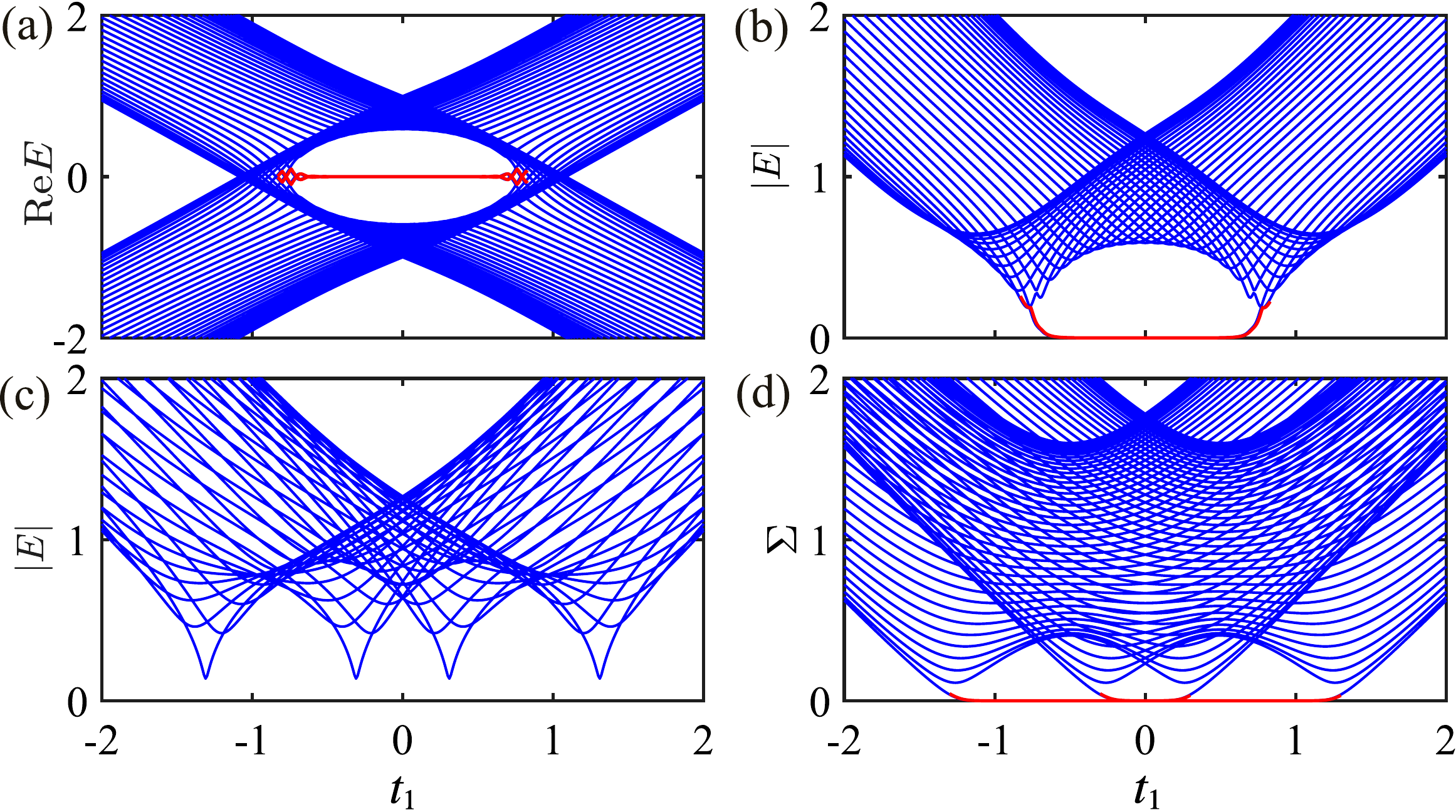}	
 	\caption{(a) Real part and (b) absolute value of the energy spectra of $H_5$ with open boundary conditions. For $|t_1|<\sqrt{ \Gamma^2-\kappa^2+(t_2^4-4\Gamma^2\kappa^2)^{1/2}}=0.83$, the system is topologically nontrivial, supporting zero-energy boundary modes (red curves). The number of unit cells is 80 with $t_2=1$, $\kappa=0.6$, and $\Gamma=0.5$.  (c) Absolute value of the energy spectrum of $H_5$ with periodic boundary conditions, where there exists four topological phase transitions. The energy spectrum is distinct from that with open boundary conditions, indicating that the mixed non-Hermitian Hamiltonian is sensitive to its boundary conditions.  (d) The singular values of $H_5$ with the open boundary conditions. According to (b,d), the singular values cannot directly reflect the topological phases of mixed non-Hermitian systems. }
 	\label{fig_h3}
 \end{figure*}

\subsection{Disordered Kitaev chain}
As an another concrete non-Hermitian model with CM, we consider a disordered Kitaev chain.  The Hamiltonian is $H_4 = H_0 +H_{\textrm{dis}}$, where
\begin{eqnarray} \label{Hpw}\nonumber
&&H_0 = \sum_n [(-tc_{n+1}^\dagger c_{n} +\Delta c_{n+1}^\dagger c_{n}^\dagger+h.c.)+\mu c_{n}^\dagger c_{n}],\\
&&H_{\textrm{dis}} =\sum_n U_n c_{n}^\dagger c_{n},
\end{eqnarray}
where $H_{\textrm{dis}}$ denotes diagonal disorder.
Here, $t$ and $\Delta$ are real numbers representing the hopping strength and the superconducting gap, respectively, and $U_n\in [-V_0/2,V_0/2]$ is the disorder satisfying an uniform distribution. As shown in Refs.~\cite{Kozii,Papaj,Shen2}, we can construct an effective non-Hermitian Hamiltonian by considering the retarded Green's function
\begin{eqnarray}
G_{\text{ret}}(k,\omega) =[\omega-H_{\text{eff}}(k)]^{-1},
\end{eqnarray}
where
\begin{eqnarray}
H_{\text{eff}} (k)\equiv H_0(k)+\Sigma_{\text{ret}}(k,\omega),
\end{eqnarray}
and $\Sigma_{\text{ret}}(k,\omega)$
is the retarded self-energy of the disorder scattering. The effective Hamiltonian has the form
\begin{equation} \label{hef}
H_{\text{eff}} (k)= -2\Delta \sin k \sigma_y + (\bar{\mu} + i\Gamma- 2t\cos k)\sigma_z,
\end{equation}
where  $\bar{\mu}$ is a renormalized chemical potential (see Appendix D). This non-Hermitian Hamiltonian effectively describes the disordered
Kitaev chain, where the disorder scattering makes the particles possess a finite lifetime, and thus broadens the bands and
shrinks the gap \cite{Kozii}. It is thus inferred that the non-Hermiticity in $H_{\text{eff}} (k)$ is attributed to the CM. The
conventional bulk-boundary correspondence holds for $H_{\text{eff}} (k)$.  The energy spectra of $H_{\text{eff}} (k)$ is shown
in Fig.~\ref{fig_kc}(a-c). When
\begin{eqnarray}
\bar{\mu}^2 < t^2[4-(\Gamma/\Delta)^2],
\end{eqnarray}
the system is topologically nontrivial supporting
Majorana zero modes (see red curves). In particular, according to Ref.~\cite{Brouwer2} and Appendix D, for $\mu=0$ and $t=1$, the
self-energy is $iV_0^2(\openone+\sigma_z)/48$. The disordered Kitaev chain is topologically nontrivial for
\begin{eqnarray}
\Delta|> V_0^2/96,
\end{eqnarray}
which agrees with the results in Refs.~\cite{Brouwer2,Pientka,DeGottardi}.  In addition, according to Fig.~\ref{fig_kc}(c,d), the
singular values of $ H_{\text{eff}} (k) $ directly reflect the topological phase of  non-Hermitian systems with CM.

\section{Non-Hermiticity with mixed LSV and CM}\label{Non-Hermiticity with mixed LSV and CM}
As discussed above, the non-Hermiticity can mainly result from LSV or CM. However, for a general non-Hermitian system, there exists \textit{both} LSV and CM, dubbed here as mixed non-Hermiticity. In this case, due to LSV, the non-Hermitian system with mixed LSV and CM is sensitive to boundary conditions. Therefore, we need to utilize the non-compact $U(1)$ gauge transformation to map such mixed non-Hermitian Hamiltonian to the one containing only a CM. Its topological phases are then determined by this transformed Hamiltonian.

We now consider a mixed non-Hermitian SSH model:
\begin{align} \label{nhssh3} \nonumber
H_5 = &\sum_n \left[(t_1+i\kappa+\Gamma)c_{\text A,n}^\dagger c_{\text B,n}+(t_1+i\kappa-\Gamma)c_{\text B,n}^\dagger c_{\text A,n}\right. \\
&~~~~~~\left.+ t_2c_{\text A,n+1}^\dagger c_{\text B,n}+t_2c_{\text B,n}^\dagger c_{\text A,n+1}\right].
\end{align}
By a non-compact $U(1)$ gauge transformation, we can map it to a non-Hermitian Hamiltonian with the only CM term left as
\begin{align}
\tilde{H}_5 = e^{-S} H_5 e^S = & \sum_n \tilde{t}_1(c_{A,n}^\dagger c_{B,n}+c_{B,n}^\dagger c_{A,n}) \nonumber \\
& +\tilde{t}_2(c_{A,n+1}^\dagger c_{B,n}+c_{B,n}^\dagger c_{A,n+1}),
\end{align}
where~\cite{Yao}
\begin{eqnarray}
e^S  =  \text{diag}(1,r,r,r^2,r^2,...,r^{L-1},r^{L-1},r^{L}),
\end{eqnarray}
and
\begin{eqnarray}
r = \sqrt{(t_1+i\kappa-\Gamma)/(t_1+i\kappa+\Gamma)},
\end{eqnarray}
Note that the Hamiltonians $\tilde{H}_5$ and $H_5$ are topologically equivalent. The effective intracell and intercell hopping strengths for the transformed Hamiltonian $\tilde{H}_5$ is obtained as
\begin{eqnarray} \nonumber
\tilde{t}_1 = \sqrt{(t_1+i\kappa-\Gamma)(t_1+i\kappa+\Gamma)} \ \ \ \text{and}\ \ \  \tilde t_2 =t_2,\\
\end{eqnarray}
respectively. Therefore, for
\begin{eqnarray}
t_1^2\leqslant \Gamma ^2-\kappa^2+\sqrt{t_2^4-4\Gamma^2\kappa^2},
\end{eqnarray}
$H_5$ is topologically nontrivial supporting zero-energy boundary modes, as shown in Fig.~\ref{fig_h3}(a,b). The energy spectrum of the mixed non-Hermitian SSH model is sensitive to its boundary conditions [see Fig.~\ref{fig_h3}(b,c)]. In addition, by comparing Fig.~\ref{fig_h3}(c) with~\ref{fig_h3}(d), the singular values of the SVD cannot directly reflect the topological phase of the mixed non-Hermitian systems.

\section{Conclusion}\label{Conclusion}
The topological phases of non-Hermitian systems are studied from the Dirac equation. Using both Dirac and current-conservation equations, we identified and investigated two classes of non-Hermiticities due to LSV and CM. We also addressed the mechanisms to break the conventional bulk-boundary correspondence in these two types of non-Hermitian systems. There exists intrinsic current and non-Hermitian skin effects in the non-Hermitian system with LSV. The bulk-boundary correspondence can be restored by mapping it to the Hermitian case via a non-compact $U(1)$ gauge transformation. In contrast, the non-Hermitian system with CM shows no intrinsic current, while the particles in this system have a finite lifetime. Moreover, there exists no non-Hermitian skin effect, and the conventional bulk-boundary correspondence holds in this system. In addition, the singular values of the SVD can be utilized to directly reflect the topological phase of this system. We also studied a general non-Hermitian system containing both LSV and CM non-Hermiticities, and suggested the approaches to generalize its bulk-boundary correspondence.

\begin{acknowledgements}
Y.R.Z. was supported by China Postdoctoral Science
Foundation (Grant No.~2018M640055) and NSFC (Grant No.~U1530401).
T.L. was supported by JSPS Postdoctoral Fellowship
(P18023). S.W.L. was supported by the research startup foundation of Dalian Maritime University in 2019 and 
the Fundamental Research Funds for the Central Universities (Grant No.~017192608). H.F. was supported by NSFC (Grant No.~11774406),
National Key R \& D Program of China (Grant Nos.~2016YFA0302104, 2016YFA0300600),
Strategic Priority Research Program of Chinese Academy of
Sciences (Grant No.~XDB28000000), and Beijing Science
Foundation (Grant No.~Y18G07). F.N. is supported in part by the:
MURI Center for Dynamic Magneto-Optics via the
Air Force Office of Scientific Research (AFOSR) (FA9550-14-1-0040),
Army Research Office (ARO) (Grant No. Grant No. W911NF-18-1-0358),
Asian Office of Aerospace Research and Development (AOARD) (Grant No. FA2386-18-1-4045),
Japan Science and Technology Agency (JST) (via the Q-LEAP program, the ImPACT program, and the CREST Grant No. JPMJCR1676),
Japan Society for the Promotion of Science (JSPS) (JSPS-RFBR Grant No. 17-52-50023, and JSPS-FWO Grant No. VS.059.18N),
the RIKEN-AIST Challenge Research Fund, and the
John Templeton Foundation.
\end{acknowledgements}

\appendix
\section{Non-Hermitian Qi-Wu-Zhang model with Lorentz symmetry violation}
The Qi-Wu-Zhang (QWZ) model is an example of a 2D model describing a Chern insulator.
The topological phases of the non-Hermitian QWZ model have been investigated using the non-Bloch-wave method in the framwork of perturbation theory in Ref.~\cite{Yao2}. However, we can solve it exactly via a non-compact $U(1)$ gauge transformation. Here, we present the details to obtain the phase diagram of the
non-Hermitian QWZ model with the Hamiltonian shown in Eq.(\ref{hc}).
By replacing $\vec{k}$ with $\left(\vec{k}+i\vec{A}\right)$, we have

\begin{eqnarray} \label{hc2}\nonumber
\tilde{H}_2(\vec{k}) =&& [\sin (k_x+iA_x) +i\delta_x]\sigma_x+[\sin (k_y+iA_y) +i\delta_y]\sigma_y\\ [0.2cm]
&&+[m-\cos (k_x+iA_x) - \cos (k_y+iA_y)]\sigma_z.
\end{eqnarray}
The spectrum of $\tilde{H}_2(\vec{k})$ can be written as
\begin{widetext}
\begin{eqnarray} \label{e2}\nonumber
E^2 =\ && [\sin (k_x+iA_x) +i\delta_x]^2+[\sin (k_y+iA_y) +i\delta_y]^2+[m-\cos (k_x+iA_x) - \cos (k_y+iA_y)]^2\\
\nonumber
=\ &&m^2+2-\delta_x^2-\delta_y^2+2i\delta_x(\cosh A_x\cdot \sin k_x+i\sinh A_x\cdot \cos k_x) + 2i\delta_y(\cosh A_y\cdot \sin k_y+i\sinh A_y\cdot \cos k_y) \\ \nonumber
&&-2m(\cosh A_x\cdot \cos k_x-i\sinh A_x\cdot \sin k_x)-2m(\cosh A_y\cdot \cos k_y-i\sinh A_y\cdot \sin k_y)\\
&&+2(\cosh A_x\cdot \cos k_x-i\sinh A_x\cdot \sin k_x)(\cosh A_y\cdot \cos k_y-i\sinh A_y\cdot \sin k_y).
\end{eqnarray}
Now we consider the real and imaginary parts of $E^2$ as
\begin{align} \label{re}
\text{Re} E^2 =  ~ & m^2+2-\delta_x^2-\delta_y^2 -2\delta_x\sinh A_x\cdot \cos k_x-2\delta_y\sinh A_y\cdot \cos k_y
-2m\cosh A_x\cdot \cos k_x-2m\cosh A_y\cdot \cos k_y  \nonumber \\
& +2\cosh A_x\cdot\cosh A_y\cdot \cos k_x\cdot \cos k_y
-2\sinh A_x\cdot\sinh A_y\cdot \sin k_x\cdot \sin k_y,
\end{align}
\begin{align} \label{ree}
\text{Im} E^2 = ~ & 2\delta_x\cosh A_x\cdot \sin k_x+2\delta_y\cosh A_y\cdot \sin k_y +2m\sinh A_x\cdot \sin k_x+2m\sinh A_y\cdot \sin k_y  \nonumber \\
& - 2\cosh A_x\cdot\sinh A_y\cdot \cos k_x\cdot\sin k_y -2\sinh A_x\cdot\cosh A_y\cdot \sin k_x\cdot \cos k_y.
\end{align}
By letting $\text{Im} E^2 = 0$, we have
\begin{eqnarray} \label{c1}
\delta_x\cosh A_x+ m\sinh A_x  -\sinh A_x\cdot\cosh A_y\cdot  \cos k_y=0,
\end{eqnarray}
\begin{eqnarray} \label{c11}
\delta_y\cosh A_y+m\sinh A_y-\cosh A_x\cdot\sinh A_y\cdot \cos k_x=0.
\end{eqnarray}
\end{widetext}
	
Firstly, we consider the special case, i.e., $\delta_x=\delta_y=\delta$. According to Ref.~\cite{Yao2}, we know that the gappless bands only appear at  high-symmetry points of the Brillouin zone, i.e., $(k_x,k_y)=(0,0)$, $(0,\pi)$, $(\pi,0)$
and $(\pi,\pi)$. For $(k_x,k_y)=(0,0)$, according to the symmetry of Eq.~(\ref{c1}) and (\ref{c11}), we can obtain
$A_x=A_y=A$, and $A$ satisfies
\begin{eqnarray} \label{c2}
&& \delta\cosh A+ m\sinh A =\sinh A\cdot\cosh A.
\end{eqnarray}
The real part of $E^2$ can be simplified as
\begin{eqnarray} \label{re1}\nonumber
\text{Re} E^2 =   m^2+2-2\delta^2 -4\delta\sinh A
-4m\cosh A+2\cosh ^2A.\\
\end{eqnarray}
Given $\text{Re} E^2 =0$, we can solve the Eqs.~(\ref{c2}) and (\ref{re1}), and the critical point is obtained as
\begin{eqnarray} \label{cp}
m = 2\sqrt{\delta^2+1}.
\end{eqnarray}
Similarly, for $(k_x,k_y)=(0,\pi)=(\pi,0)$, the corresponding critical point is $m =0$, while for $(k_x,k_y)=(\pi,\pi)$, the critical point is
\begin{eqnarray}
m = -2\sqrt{\delta^2+1}.
\end{eqnarray}
For the general case $\delta_x \neq \delta_y$, we can apply the same procedures at the high-symmetry points $(k_x,k_y)=(0,0)$, $(0,\pi)$, $(\pi,0)$ and $(\pi,\pi)$. For $(k_x,k_y)=(0,0),(\pi,\pi)$, the critical points can be solved as
\begin{eqnarray}
m = \pm(\sqrt{1+\delta_x^2}+\sqrt{1+\delta_y^2}),
\end{eqnarray}
and other two critical points are
\begin{eqnarray}
m = \pm(\sqrt{1+\delta_x^2}-\sqrt{1+\delta_y^2}).
\end{eqnarray}

\section{Geometric description of topological phase transitions of non-Hermitian systems with complex mass}

Here, based on quantum field theory, we give a geometric description of topological phase transitions in Hermitian systems,  and then generalize it to the non-Hermitian case with complex mass. This provides a new viewpoint to understand the topological properties of non-Hermitian systems.

In quantum field theory, the 4D Dirac spinor is a representation of the 4D Clifford algebra.
The generators of the algebra are $\gamma^{\mu}$ matrices, which satisfy
\begin{equation}
\{ \gamma^{\mu},\gamma^{\nu}\} =2\eta^{\mu\nu}.
\end{equation}
It mathematically turns out that the 4D Dirac representation of
the Lorentz group is reducible due to $SO(4)\simeq SU_{L}(2)\times SU_{R}(2)$.
Hence we can define the chirality with the operator $P_{L,R}={(1\mp\gamma_{*})}/{2}$, and $\psi_{L,R}=P_{L,R}\psi$. This forms a 2D representation
by writing the Dirac spinor as
\begin{equation}
\psi=\left(\begin{array}{c}
\psi_{L}\\
\psi_{R}
\end{array}\right).
\end{equation}
Here, $\psi_{L,R}$ transform under $SU_{L,R}(2)$, respectively,
and is named as left- and right-handed Weyl spinor. Note that
$\gamma_{*}$ is defined as $\gamma_{*}=i\gamma^{0}\gamma^{1}\gamma^{2}\gamma^{3}$,
and $\psi_{L,R}$ are the eigenstates of $\gamma_{*}$. Then let us
accordingly consider a global transformation generated by $\gamma_{*}$, i.e.,
the chiral transformation $\psi\rightarrow e^{i\theta\gamma_{*}}\psi$.
Analyzing its infinitesimal transformation, we can obtain the corresponding
Noether current $j_{*}^{\mu}=\bar{\psi}\gamma^{\mu}\gamma_{*}\psi$,
which satisfies
\begin{equation}
\partial_{\mu}j_{*}^{\mu}=2im\bar{\psi}\gamma_{*}\psi.
\end{equation}
Note that $j_{*}^{\mu}$ is conserved only if $m=0$, indicating that a
chiral phase transition occurs at $m=0$.

Alternatively, there exists a geometric interpretation of the topological
or chiral phase transition in fermionic systems. The phase transition
could be identified as a Dirac field defined in a space-time with different
topologies. Without loss of generality, let us investigate the dynamics
of a massless spinor $\psi(X)$ in $D+1$ dimensional Minkowskian
spacetime $\mathbb{R}^{D+1}$ parameterized by $X\equiv\{ X^{M}\} =\{ x^{\mu},x^{D+1}\equiv y\} $,
and the index $\mu$ runs from $0,1,\cdots,D-1$. The Dirac equation for
the spinor is given as
\begin{equation}
\gamma^{M}\partial_{M}\psi(X)=0,\label{eq:4}
\end{equation}
where the matrices $\gamma^{M}\equiv\{ \gamma^{\mu},\Gamma\} $
satisfy the $(D+1)$ dimensional Clifford algebra $\{ \gamma^{M},\gamma^{N}\} =2\eta^{MN}$.
In order to take into account a mass term, we could compactify one spatial
direction, denoted by $y$, of the space-time on a circle $S^{1}$
with radius $L$ so that $\tau(y)=\tau(y+2\pi L)$.
Here $\tau$ denotes the line element on $S^{1}$ and the topology
of the spacetime now becomes $\mathbb{R}^{D}\times S^{1}$. Then, we expand the spinor by its Fourier modes as
\begin{equation}
\psi(X)=\sum_{n}e^{iky}\Psi_{n}(x),\label{eq:5}
\end{equation}
where $k$ is the momentum on the $y$ direction. Note that the boundary condition
could be either periodic or anti-periodic, i.e., $\psi(x,\tau)=\pm\psi(x,\tau+2\pi)$
for a spinor, and we consider the anti-periodic condition here, since
a fermionic field is not observable. Therefore, $k$ has to satisfy
the quantization condition
\begin{equation}
k=\frac{l}{2L},\hspace{0.2 in}l\in Z.
\end{equation}
By plugging (\ref{eq:5}) into (\ref{eq:4}), we obtain
\begin{align}
\gamma^{M}\partial_{M}\psi(X)
=  \sum_{n}(\gamma^{\mu}\partial_{\mu}+\Gamma\partial_{y})e^{iky}\Psi_{n}(x)=0,
\end{align}
which leads to
\begin{equation}
(i\gamma^{\mu}\partial_{\mu}-k\Gamma)\Psi_{n}(x)=0.\label{eq:8}
\end{equation}
Hence we can define the effective mass
\begin{eqnarray}
m_{l}^{2}\equiv|{l}\Gamma/{(2L)}|^{2}={l^{2}}/{(4L^{2})}
\end{eqnarray}
to obtain the massive Dirac equation in $D$ dimension. As discussed here,
the Dirac equation is chirally symmetric only for the massless spinor,
so the chiral transition can be geometrically interpreted as going from $L\rightarrow\infty$
to a finite $L$. The former corresponds to a space-time with topology
$\mathbb{R}^{D+1}$, and the latter corresponds to its topology $\mathbb{R}^{D}\times S^{1}$.
Thus the chiral transition could be understood as a topological
transition of the $(D+1)$ dimensional spacetime from $\mathbb{R}^{D+1}$
to $\mathbb{R}^{D}\times S^{1}$. Furthermore, the equation (\ref{eq:8}) also
leads to an effective vertex $k\bar{\Psi}\Gamma\Psi$ in the associated
action. If we accordingly compare Eq.~(\ref{eq:8}) with the coupling
term in the gauge theory involving fermions, i.e., $\bar{\Psi}\gamma^{\mu}A_{\mu}\Psi$,
the effective vertex $k\bar{\Psi}\Gamma\Psi$ can be obtained by
treating $A_{\mu}$ as an external field. In this sense, by taking
into account  the massive Dirac equation
\begin{eqnarray}
(i\gamma^{\mu}\partial_{\mu}-m)\Psi(x)=0,
\end{eqnarray}
we can define an imaginary mass,
\begin{eqnarray}
(m^{\textrm{i}})^{2}\equiv[{l}\Gamma/({2L})]^{2}=-{l^{2}}/({4L^{2}}),
\end{eqnarray}
in order to describe such an effective interaction in $D$-dimensional
space-time.

\section{Minimal coupling and singular value decomposition}
The minimal coupling \cite{Ueda} and the singular value decomposition (SVD) can be used to explore the topological properties of the non-Hermitian systems with complex mass. In addition, the SVD can overcome the precision problem of numerically diagonalizing a non-Hermitian Hamiltonian.  In this section, we demonstrate that the minimal coupling is equivalent to calculating the singular value decomposition for the non-Hermitian Hamiltonian with complex mass. We write the minimal coupling Hamiltonian
\begin{eqnarray}
H_{cp} = \mathcal{H}_{\text{CM}}\otimes\sigma_{+}+\mathcal{H}^\dagger_{\text{CM}}\otimes\sigma_{-}
\end{eqnarray}
as
\begin{eqnarray} \label{H2}
H_{cp} = \left(\begin{matrix}
0 & \mathcal{H}_{\text{CM}} \\
\mathcal{H}^\dagger_{\text{CM}} & 0
\end{matrix}\right).
\end{eqnarray}
$\mathcal{H}_{\text{CM}}$ satisfies the singular value decomposition as
\begin{eqnarray}\nonumber
\mathcal{H}_{\text{CM}}=U\varSigma V^\dagger,\\
\mathcal{H}^\dagger_{\text{CM}}=V\varSigma U^\dagger,
\end{eqnarray}
where $\varSigma$ is a positive diagonal matrix. Thus, we have
\begin{eqnarray} \label{sig}
H_{cp} = \left(\begin{matrix}
0 & U \\
V & 0
\end{matrix}\right)
\left(\begin{matrix}
0 & \varSigma \\
\varSigma & 0
\end{matrix}\right)
\left(\begin{matrix}
0 & V^\dagger \\
U^\dagger & 0
\end{matrix}\right),
\end{eqnarray}
with
\begin{eqnarray} \label{H3}
\left(\begin{matrix}
0 & \varSigma \\
\varSigma & 0
\end{matrix}\right)=
\frac{1}{\sqrt{2}}\left(\begin{matrix}
1 & 1 \\\nonumber
-1 & 1
\end{matrix}\right)\cdot
\left(\begin{matrix}
-\varSigma & 0 \\\
0 & \varSigma
\end{matrix}\right)\cdot
\frac{1}{\sqrt{2}}\left(\begin{matrix}
1 & -1 \\
1 & 1
\end{matrix}\right).\\
\end{eqnarray}
The Hermitian Hamiltonian $H_{cp}$ can be diagonalized as
\begin{eqnarray} \label{H4}
H_{cp} =
\Lambda
\left(\begin{matrix}
-\varSigma & 0 \\
0 & \varSigma
\end{matrix}\right)\Lambda^\dagger, \hspace{0.2in}
\end{eqnarray}
with
\begin{eqnarray}
\Lambda =
\left(\begin{matrix}
0 & U \\
V & 0
\end{matrix}\right)\cdot
\frac{1}{\sqrt{2}}\left(\begin{matrix}
1 & 1 \\
-1 & 1
\end{matrix}\right)\cdot
\end{eqnarray}
It is clear that the eigenvalues and eigenvectors of $H$ are the singular values and singular matrix of $\mathcal{H}_{\text{CM}}$, respectively. Therefore, the minimal coupling is equivalent to solving the singular value decomposition of the non-Hermitian system with complex mass.

\section{Self-energy of disorder Kitaev chain}
To obtain the effective Hamiltonian of the disordered Kitaev chain, we need to calculate the self-energy of the disorder scattering.
Here, we derive the form of the self-energy  using  Feynman diagrams.
The Hamiltonian of the disordered Kitaev chain has the form
\begin{eqnarray} \label{hk} \nonumber
H_4 &&= \sum_k[\psi_k^\dagger[-2\Delta \sin k \sigma_y + (\mu- 2t\cos k)\sigma_z]\psi_k \\
&&+\sum_q V(q)\psi_{k+q}^\dagger\varLambda\psi_k],
\end{eqnarray}
where $\psi_k=(c_k,c_{-k}^\dagger)^T$, $V(q)=\sum_n U_ne^{-inq}$, and $\varLambda=(\openone+\sigma_z)/2$.
In addition, $V(q)$ satisfies
\begin{eqnarray} \label{disave} \nonumber
\langle V(q)\rangle = 0,   \ \ \ \ \ \ \langle V(q_1)V(q_2)\rangle = \frac{V_0^2}{12}\delta(q_1+q_2),\\
\end{eqnarray}
where $\langle...\rangle$ means the average over disorder, and $\delta(\cdot)$ is the Dirac delta function.

We consider the Matsubara self-energy of non-crossing diagrams, and the $n$th-order Matsubara self-energy $\Sigma_n(k,i\omega)$ has the form
\begin{eqnarray} \label{se} \nonumber
&&\Sigma_1(k,i\omega)=\langle V(0)\rangle = 0, \\[0.2cm] \nonumber
&&\Sigma_2(k,i\omega)= \sum_q \langle V(k-q)V(q-k)\rangle\varLambda G_0(q,i\omega)\varLambda\\
&& ...
\end{eqnarray}
with
\begin{eqnarray} \label{g0}
G_0(q,i\omega)=\frac{1}{i\omega-H_0(q)},
\end{eqnarray}
where $H_0(q)$ is the free part of $H_4$.
Thus, the total Matsubara self-energy reads
\begin{eqnarray}  \nonumber
\Sigma(k,i\omega)=g(k,i\omega)\cdot\varLambda=g(k,i\omega)/2+g(k,i\omega)/2\cdot\sigma_z, \\
\end{eqnarray}
  where $g(k,i\omega)$ is a $\left(k, \omega\right)$-dependent function. By analytic continuation $i\omega\rightarrow\omega+i0^+$, we can obtain the retarded self-energy
  \begin{eqnarray} \nonumber
  \Sigma_{\text{ret}}(k,\omega)&&=\Sigma(k,\omega+i0^+)\\ \nonumber
  &&=g(k,\omega+i0^+)/2+g(k,\omega+i0^+)/2\cdot\sigma_z.\\
  \end{eqnarray}
The real part of $g(k,\omega+i0^+)$ renormalizes the chemical potential, while the imaginary part gives the finite lifetimes of the particles. Therefore, the effective Hamiltonian of $H_4$ reads
\begin{equation} \label{heff}
H_{\text{eff}} (k)= -2\Delta \sin k \sigma_y + (\bar{\mu} + i\Gamma- 2t\cos k)\sigma_z,
\end{equation}
where
\begin{eqnarray}  \nonumber
\bar{\mu}&&=\mu+\frac{1}{2}\text{Re}[g(k,\omega+i0^+)], \\
\Gamma&& = \frac{1}{2}\text{Im}[g(k,\omega+i0^+)]/2.
\end{eqnarray}


%

\end{document}